\documentclass[fleqn,usenatbib]{mnras}

\usepackage{txfonts}

\usepackage[T1]{fontenc}

\DeclareRobustCommand{\VAN}[3]{#2}
\let\VANthebibliography\thebibliography
\def\thebibliography{\DeclareRobustCommand{\VAN}[3]{##3}\VANthebibliography}

\usepackage{graphicx}
\usepackage{xcolor,ulem}	% Including figure files
\usepackage{amssymb}	% Extra maths symbols

\title[$z \sim 1$ Cluster Properties in HR5]{Star-formation Properties of $z \sim 1$ Galaxy Clusters and Groups from Horizon Run  5 }

\author[S.-K. Lee et al.]{
Seong-Kook Lee,$^{1,2}$
Changbom Park,$^{3}$\thanks{E-mail: cbp@kias.re.kr}
Juhan Kim$^{3}$\thanks{E-mail: kjhan@kias.re.kr}
Jaehyun Lee$^{3,4}$ 
Brad K. Gibson,$^{5}$
\newauthor
Yonghwi Kim,$^{6}$
and
C. Gareth Few$^{5}$
\\
$^{1}$SNU Astronomy Research Center, Department of Physics and Astronomy, Seoul National Universty, Seoul, Korea\\
$^{2}$Astronomy Program, Department of Physics and Astronomy, Seoul National Universty, Seoul, Korea\\
$^{3}$Korea Institute for Advanced Study, Seoul, Korea\\
$^{4}$Korea Astronomy and Space Science Institute, Daejeon, Korea\\
$^{5}$E.A. Milne Centre for Astrophysics, University of Hull, Hull, HU6 7RX, United Kingdom\\
$^{6}$Korea Institute of Science and Technology Information, 245 Daehak-ro, Yuseong-gu, Daejeon, 34141, Korea\\
}

\date{Accepted XXX. Received YYY; in original form ZZZ}

\pubyear{2023}

\begin{document}
\label{firstpage}
\pagerange{\pageref{firstpage}--\pageref{lastpage}}
\maketitle

\begin{abstract}
Quiescent galaxies are predominantly observed in local galaxy clusters. However, the fraction of quiescent galaxies in high-redshift clusters significantly 
varies among different clusters. 
In this study, we present the results of an analysis of the star formation properties of $z \sim 0.87$ clusters and groups from a cosmological 
hydrodynamical simulation Horizon Run 5. 
We investigate the correlation between the quiescent galaxy fraction (QF) of these model clusters/groups and their various internal or external properties.  
We find that halo mass is one of the most important characteristics as higher-mass clusters and groups have higher QFs. 
We also find that other properties such as stellar-mass ratio and FoF fraction, measures the proportion of the area around a cluster occupied by dense structures,
may mildly affect the QFs of clusters and groups. 
This may indicate that the evolutionary history as well as the large-scale environment of clusters and groups also play a certain role in determining 
the star formation status of high-redshift galaxy clusters and groups. 
\end{abstract}

\begin{keywords}
galaxies: clusters: general -- large-scale structure of Universe -- methods: numerical -- galaxies: evolution
\end{keywords}

%%%%%%%%%%%%%%%%%%%%%%%%%%%%%%%%%%%%%%%%%%%%%%%%%%

%%%%%%%%%%%%%%%%% BODY OF PAPER %%%%%%%%%%%%%%%%%%

\section{Introduction}

Galaxy clusters provide interesting laboratory to study the evolution of galaxies. 
There are many interesting issues regarding the evolution of galaxies and galaxy clusters in these dense structures. 
%%\kjhanadd{(PLEASE SPECIFY SEVERAL ISSUES HERE AND PROVIDE REFERENCES)} 
One of these issues is that how the  star formation of galaxies would evolve in dense environment, such as galaxy clusters or groups.

As it has usually been understood, the regulation of star-formation (SF) activities of galaxies has two competing 
origins: nature \citep[e.g.][]{kau03,jim05,bal06} versus nurture \citep[e.g.][]{lew02,kau04,bal06}
 (or, in other words, internal versus external effects). 
Therefore, it would be important to quantitatively study the correlation of the SF activities of cluster galaxies 
with their mass and environments.

Lots of observational results indicate that, at high redshift ($z>1$), the stellar-mass dependence on the SF activities is generally 
greater than the environmental effects\citep[e.g.][]{pen10,lee15,cha20,gu21}. 
At $z>1$, more massive galaxies stopped their star-formation earlier on average, 
and therefore the fraction of quenched galaxies shows strong dependence on stellar mass 
\citep[e.g.][]{pen10,lee15}.
As time goes by, the environmental effects such as the ram-pressure stripping \citep{gun72, aba99}, 
strangulation \citep{lar80,bal00, pen15}, and galaxy harassment \citep{moo96} becomes dominant 
suppressing the SF activities and changing the related physical properties of cluster galaxies. 

However, during this transition era (or at the cosmic noon), the SF quenching does not occur at the same pace 
among galaxy clusters.
In other words, there exists a cluster-to-cluster variation in the status of SF quenching 
(or quantitatively the quiescent galaxy fraction) among clusters at similar redshift 
\citep{lee15,alb16,hay19,lee19}. 

Therefore, an interesting question may be raised on the possible origin of the variation of the quiescent galaxy fraction. 
Already, \citet{lee19} have shown that the observed fraction of quiescent galaxies in a cluster is dependent of the 
large-scale environments of galaxy clusters having a lower value for a denser cluster environments.  
As a possible explanation for this observed results, they proposed a `Web-feeding model' 
\citep{lee19} which supposes that star-forming galaxies infalling into the cluster through the local cosmic web structures 
keep low the quiescent galaxy fraction for some period of time. \citep[e.g.][]{ell01,tra05,lub09}.

Halo mass can also be a crucial factor determining the star formation characteristics of galaxy clusters.
Previous studies on this matter are still under debate. 
Some suggests that the star formation characteristics of galaxy clusters vary with the halo-mass of the clusters 
\citep[e.g.][]{bla07,kim09,wet12,woo13}, 
while other studies yield results indicating that there is no correlation between the star formation properties of 
galaxy clusters and the halo-mass of the clusters \citep[e.g.][]{fin08,van08,pas09}.
And certain environmental dependence of star formation properties at the cluster scale 
may be interpreted in terms of halo-mass dependence \citep[e.g.][]{haa12}.

Also, there would be other properties or parameters of galaxy clusters which may have some correlations with 
the quiescent galaxy fraction of clusters and this is the main focus in this study. 
For this purpose, we use model galaxies and galaxy clusters from Horizon Run 5 (HR5) simulation. 
Several studies have investigated the evolution of star formation in cluster environments through 
various simulations.
\citet{bah17} explored the influence of the cluster environment on galaxy formation through 
the use of the Hydrangea simulations, which comprise a set of cosmological hydrodynamical 
zoom-in simulations.
\citet{rhe20} investigated the quenching history of galaxies, exploring both its environmental 
and mass dependence through the use of the Yonsei Zoom-in Cluster Simulation (YZiCS).
\citet{jeo22} examined the star-formation history of galaxies through the Horizon-AGN simulation, 
discovering the dependence of the star-formation decay time-scale on both the stellar mass and 
environments of galaxies.
Our study can be viewed as an extension of these and other previous studies in this line of research.
However, our analysis goes beyond, encompassing the influence of cosmic large-scale structures on 
the star formation characteristics of galaxies within clusters, operating on a larger scale than the 
individual sizes of each galaxy cluster.

In Section \ref{sec:dat}, we explain the simulation, as well as sample selection 
procedure. Then we present the results of analysis about the effects of various 
properties on the star-formation properties of galaxy clusters in Section \ref{sec:results} 
with discussion in Section~\ref{sec:disc}.
We summarize our findings in Section \ref{sec:conc}.
%We adopt standard cosmology with ($\Omega_{m}$, $\Omega_{\Lambda}$) = (0.3, 0.7) and 
%$H_{0}$ = 70 km s$^{-1}$ Mpc$^{-1}$, which is supported by observations in the past decades 
%\citep[e.g.][]{im97}.

\section{Simulation and Sample}
\label{sec:dat}

\subsection{Horizon Run 5 Simulation: Zoom-in Simulation}
\label{sec:hr5sim}

HR5 is a cosmological hydrodynamical zoom-in simulation aiming at covering 
a wide  range of cosmic structures in a (1048.6 cMpc)$^3$ volume, with a spatial resolution 
down to $\sim1\,$pkpc. 
HR5 is conducted using an adaptive mesh refinement code RAMSES~\citep{tey02} 
that is optimized for the OpenMP plus MPI parallelism~\citep{leej21}. 
HR5 is based on the cosmological parameters of $\Omega_{\rm m}=0.3$, 
$\Omega_{\Lambda}=0.7$, $\Omega_{\rm b}=0.047$, $\sigma_8=0.816$, 
and $h=0.684$ that are compatible with the Planck data~\citep{pla16}. 
The initial condition of HR5 is generated using the MUSIC package~\citep{hah11}, 
with a second-order Lagrangian scheme~\citep[2LPT;][]{sco98,lhu14}. 
The volume of HR5 is configured to have a high-resolution region of the cuboid geometry 
with a dimension of  $1048.6\times119.0\times127.2\,{\rm cMpc}^3$. 
The cosmological box has 256 coarse grids (level 8, $\Delta x=4.10\,$cMpc) on a side 
and the zoomed region has 8192 grids (level 13, $\Delta x=0.128\,$cMpc) along the 
major axis in the initial condition. 
The high-resolution region initially contains $8192\times930\times994$ grids and dark 
matter particles, and is surrounded by the padding grids of levels from 9 to 12. 
The dark matter particles have a mass of $6.89\times10^7\,M_\odot$ in the 
high-resolution region (level 13), and their mass increases by a factor of 8 with 
a decreasing grid level. 
The grids are adaptively refined down to $\Delta x\sim1\,$kpc when their density 
exceeds eight times the dark matter particle mass at level 13. 
HR5 completes at $z=0.625$.

HR5 is conducted by utilizing the physical processes governing the evolution of 
baryonic components that are equipped in subgrid forms in RAMSES. Gas 
cooling rates are computed using the cooling functions of \citet{sut93} 
in a temperature range of $10^4-10^{8.5}\,$K and fine-structure line cooling 
is computed down to $\sim750\,$K using the cooling rates of \citet{dal72}. 
Cosmic reionization is approximated by assuming a uniform UV background~\citep{haa96}. 
A star formation rate is computed by utilizing the statistical approach of \citet{ras06}. 
A stellar particle has a minimum mass of $2.56\times10^6\,M_\odot$ and its mass 
evolves over time based on a Chabrier initial mass function~\citep{cha03}. 
Supernova feedback operates in thermal and kinetic modes~\citep{dub08} and
 AGN feedback operates in radio-jet and quasar modes which is switched by the 
Eddington ratio~\citep{dub12}. 
Massive black holes (MHBs) are initially seeded with a mass of $10^4$ M$_{\odot}$ 
in the grids that have gas densities higher than the threshold of star formation and 
no other neighboring MBHs within 50 kpc~\citep{dub14b}. 
MBHs increase their mass via accretion and MBH coalescence, and their angular 
momentum obtained from the mass growing processes are traced~\citep{dub14a}. 
Metal enrichment is computed using the method proposed by \citet{few12} based 
on a Chabrier initial mass function. 
The abundance of H, O, and Fe are traced individually. 
Further details of HR5 are given in \citet{leej21}.

The bound objects in HR5 are identified using the Physically Self-Bound (PSB)-based 
galaxy finder~\citep[PGalF, see Appendix A of][for details]{kim23}. 
First, PGalF finds Friends-of-Friends (FoF) objects from all the mass components of 
dark matter, gas, stars, and massive black holes by utilizing the adaptive linking length 
scheme of PGalF Self-bound sub-objects are sorted out from the FoF objects by 
measuring total energy and tidal radius from the local density peaks found in the 
coordinate-free stellar or dark matter density fields of FoF objects. 
A sub-object is defined as a galaxy when it contains sufficient stellar particles for 
constructing a stellar mass density field. We classify the mass components that 
are not gravitationally bound to any sub-objects as unbound components in a FoF object. 
The evolution of galaxies in HR5 is traced using the merger trees constructed 
based on the stellar particle membership of galaxies~\citep[see][for details]{park22,lee23}.

\subsection{Halo Sample}
\label{sec:hsamp}

From the HR5 simulation data, we select a snapshot data at redshift 
$z=0.873$. This redshift epoch was selected for the comparison with 
the observational results of \citet{lee19}. 
The snapshot catalog contains clusters and groups of halos which have member galaxies.
In this snapshot data, we sampled galaxy groups or clusters with the mass condition of 
log ($M_{200}/{\rm M_{\odot}}$) $\ge 13.0$ where $M_{200}$ is the (total) halo mass 
within the radius within which mean matter density is 200 times the critical density at a 
given redshift.

From these halos, we first exclude overlapping binary halos when the distance 
between two halos is closer than the sum of their virial radius ($R_{200}$) of each halo. 
This is because when two groups or clusters are located close enough, the star formation (SF) properties 
of a halo can be affected by the gravitational influence of a nearby halo. 
Then, with the similar reason, we further exclude from the sample the seriously asymmetric halos 
which might be in the course of halo merging. 
A projected two-dimensional asymmetry parameter (on the $x-y$ plane) is defined as 
\begin{equation}
    A_{x,y} = \frac{\Sigma_{i,j} |M(i,j) - M_{180}(i,j)|}{\Sigma_{i,j} |M(i,j)|}, 
\label{eq:asymm}
\end{equation}
where $M(i,j)$ and $M_{180}(i,j)$ are the mass densities of a pixel ($i, j$) before and after 
180 degree rotation around the center of the halo on the $x-y$ plane,  respectively. 
We apply the asymmetry cut of $\mathrm{MAX} (A) \ge 0.8$ to reduce a possible contamination 
to exclude the unvirialized cases of clusters and groups.
Here, a $\mathrm{MAX} (A)$ is the maximum value among $A_{x,y}$, $A_{y,z}$, and $A_{z,x}$. 
We also exclude galaxy groups with fewer than 10 member galaxies from the sample.

After the trimming-down procedure, we finally have 224 groups and clusters, 
which will be used for our analysis in this study.

\section{Results}
\label{sec:results}

\subsection{Star Formation Properties of Halos}
\label{sec:sfindi}

\begin{figure}
    \includegraphics[width=\columnwidth]{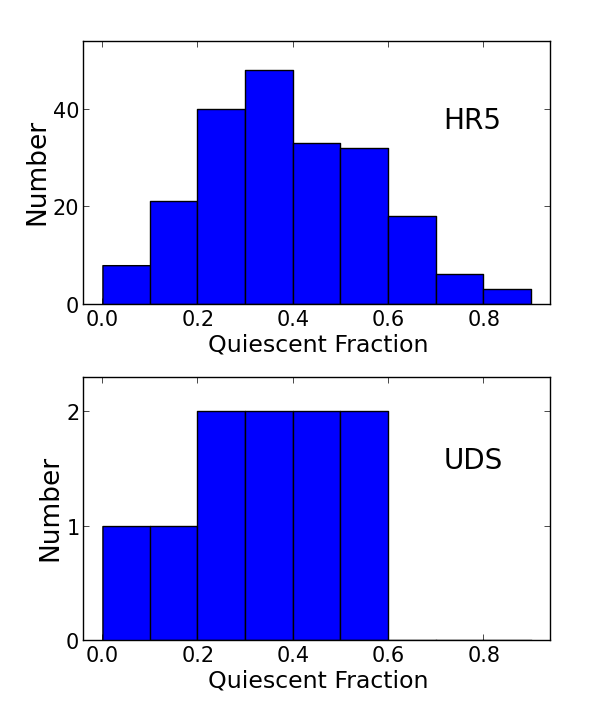}
    \caption{Distributions of QFs of galaxy groups and cluster of HR5 {\bf [top]} and of 
observed galaxy groups and clusters in the UDS {\bf [bottom]}. The QF distributions of model and 
observed galaxies are consistent.}
    \label{fig:qfdsthruds}
\end{figure}

The main aim of this work is to investigate the star formation properties of high-redshift 
galaxy groups and clusters. 
We use the quiescent galaxy fraction (QF) as a proxy for the SF activities of a cluster and group. 
Here, the QF is defined as the number fraction of quiescent galaxies among total member 
galaxies with a stellar mass of log ($M_{*}/M_{\odot}) \ge 9.1$ in each group or cluster. 
The stellar-mass cut is chosen following \citet{lee19}, ensuring a fair comparison with 
the results of \citet{lee19}.
We classify a galaxy as quiescent if the specific star-formation rate (sSFR) is 
smaller than the one third of $1/t(z)$, where $t(z)$ is the age of the Universe at target redshift $z$, 
following \citet{dam09} and \citet{lee15}. 
At $z=0.873$, when t(z)=6.43 Gyr, the sSFR cut is $5.18 \times 10^{-11}$ yr$^{-1}$.
The upper panel of Figure~\ref{fig:qfdsthruds} shows the distribution of quiescent galaxy fractions 
of our sample halos and the bottom panel shows the QF distribution of observed galaxy clusters in the 
redshift range of $0.744 \le z \le 0.919$ found in the Ultra Deep Survey (UDS) by \citet{lee15}. 
This two distributions are consistent with each other, while the number of groups/clusters in the UDS 
is much smaller mainly due to its small volume. 
We conducted a Kolmogorov-Smirnov (K-S) test on two QF distributions, resulting in a $p$-value of 0.713. 
This indicates that the likelihood of HR5 following a QF distribution different from 
the observed one is considered to be low. 

\begin{figure}
    \includegraphics[width=\columnwidth]{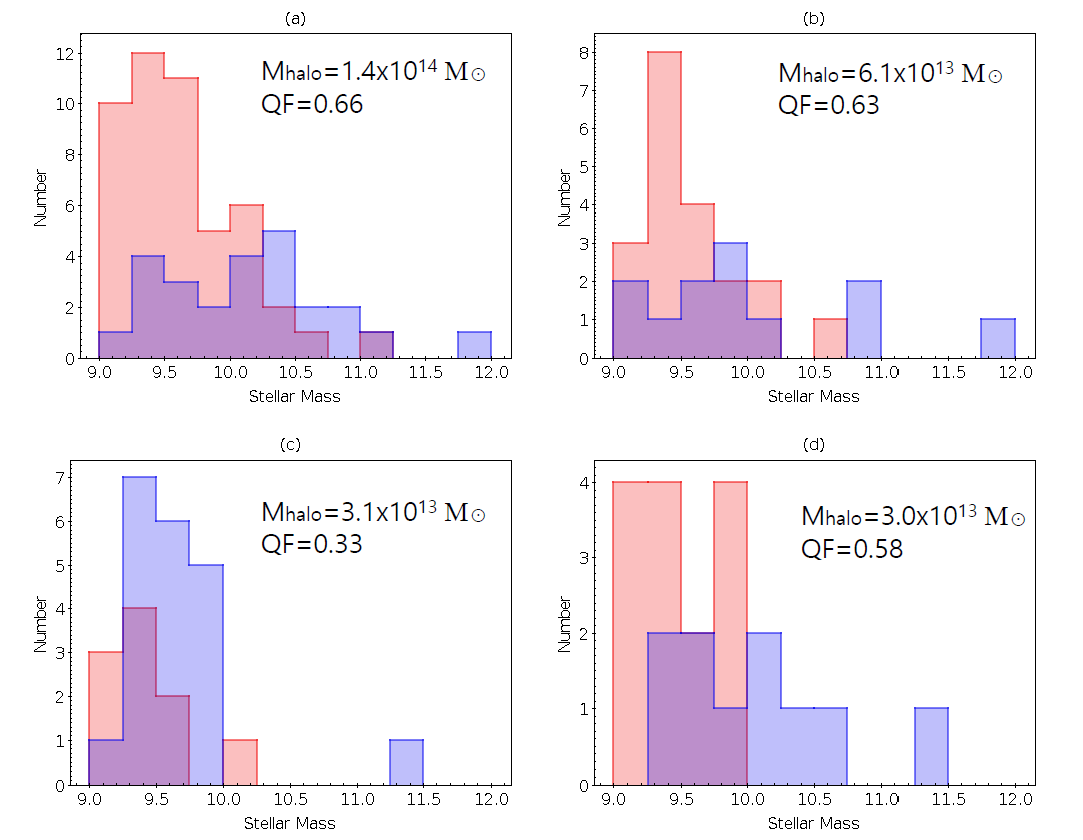}
    \caption{Stellar mass distributions of member galaxies of four examplary halos with various halo masses 
and QFs. For each halo, stellar mass distributions are shown for star-forming (blue) and 
quiescent (red) galaxies, separately. (a) A halo with halo mass, $M_{200}=1.4\times 10^{14} M_{\odot}$ and 
QF=0.66. (b) $M_{200}=6.1\times 10^{13} M_{\odot}$ and QF=0.63. (c) $M_{200}=3.1\times 10^{13} M_{\odot}$ and 
QF=0.33. (d) $M_{200}=3.0\times 10^{13} M_{\odot}$ and QF=0.58.}
    \label{fig:lmsdstex}
\end{figure}

Figure~\ref{fig:lmsdstex} shows the stellar mass distribution of member galaxies 
for several examplary halos with various halo masses and QFs. 
In this figure, the blue and red histograms show the distributions of star-forming and quiescent member 
galaxies of each halo, respectively. As seen in this figure (e.g., comparing panels (c) and (d)), 
the variation in QF among halos with similar halo masses seems to be mainly caused by the low-mass galaxies.
On the other hand, as can be seen by comparing panels (a), (b), and (d), 
halos with high QF ($> 0.55$) show similar stellar-mass distirbutions for 
star-forming and quiescent galaxies, regardless of their halo mass.
And, after we investigate the star-formation properties of member galaxies of our sample halos, 
we found that massive halos tend to have high QFs, while low-mass halos span a wider range of QF.

\subsection{Quiescent Galaxy Fraction and Halo Mass} 
\label{sec:qfhmass}

In this section, we investigate the correlation between this QF and the halo mass.
It has been known that several cluster properties show the halo-mass dependency to some degree. 
And we confirm the strong dependence of QF on the halo mass as shown below.

The upper panel of Figure~\ref{fig:qfmh} shows that there is a positive correlation between 
halo mass and QF of our sample clusters and groups: more massive halos have a higher QF. 
In this figure, the median values and their corresponding scaled MAD (Median Absolute Deviation) 
are depicted as red squares and error bars for each halo-mass bin.
The halo-mass range of  each mass bin is given as (1) $(1.0-1.5) \times 10^{13}~{\rm M_{\odot}}$, 
(2)  $(1.5-2.5) \times 10^{13}~{\rm M_{\odot}}$, (3)  $(2.5-4.0) \times 10^{13}~{\rm M_{\odot}}$, 
(4)  $(4.0-8.0) \times 10^{13}~{\rm M_{\odot}}$, (5)  $\ge 8.0 \times 10^{13}~{\rm M_{\odot}}$.

As for the correlation between halo-mass and QF, we have found the best-fit as follows: 
\begin{equation}
    {\rm QF} = 0.39 \times \log (M_{200}/{\rm M_{\odot}}) - 4.8. 
\label{eq:qfm200}
\end{equation}
This linear correlation, represented by the black line in the upper panel of Figure~\ref{fig:qfmh}, 
depicts the `halo star-formation main-sequence (HSFMS)'. 

In figure~\ref{fig:qfmh}, it is clear that there is a lower envelope of distributions, 
i.e. the minimum QF for given halo mass, with a slope of $\sim 0.5$. 
The existence of this lower envelope and the HSFMS indicate that the QF 
of high-redshift clusters largely depend on halo mass: star-formation 
quenching of cluster member galaxies first occurs for massive halos, 
and proceed toward lower-mass halos with time. 
This halo-mass dependence of star-formation quenching is consistent 
with observational results \citep[e.g.][]{kim09,woo13}.

\begin{figure}
    \includegraphics[width=\columnwidth]{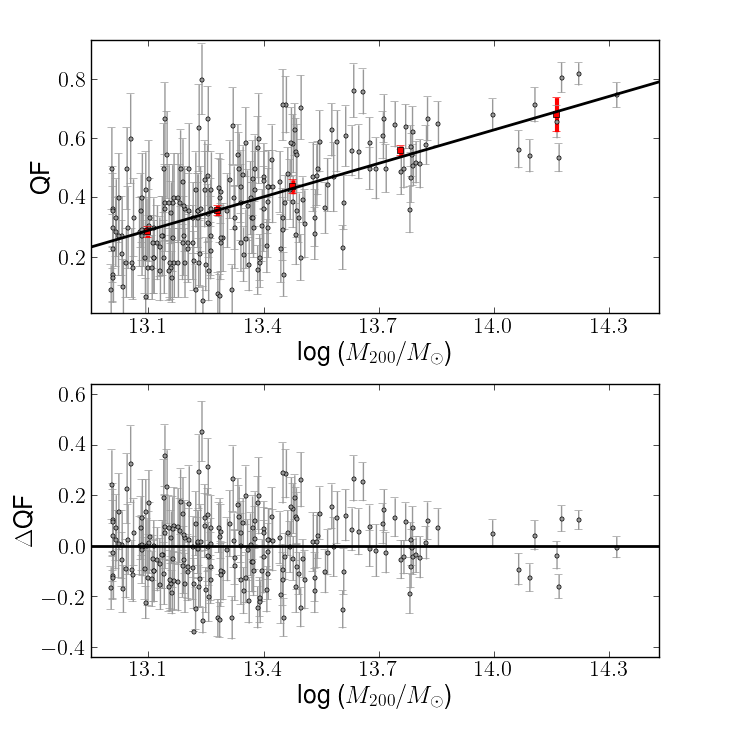}
    \caption{{\bf [Top]} Correlation between halo mass ($M_{200}$) and quiescent galaxy fraction (QF) of 
clusters/groups. Red square symbols, along with their respective error bars, denote the median and scaled 
MAD of QF within each halo mass bin. The black straight line shows the best-fit correlation between 
halo-mass and QF. {\bf [Bottom]} $M_{200}$-dependent scatter of QFs of individual halos. Here, $\Delta$QF, 
defined as (QF-QF$_{\rm best-fit}$), is the deviation from the best-fit HSFMS (Equation~\ref{eq:qfm200}) 
of indiviudal clusters/groups.}
    \label{fig:qfmh}
\end{figure}

\subsection{Quiescent Galaxy Fraction and Stellar Mass Ratio}
\label{sec:qfsmsratio}

Even with a positive correlation between the QF and the halo mass, a large scatter can 
be seen for a given halo-mass bin, especially for small-mass halos, 
which may mean that there is other hidden parametric dimensions on the QF distribution. 
In addition to halo mass, halo assembly history or evolutionary stage can be another 
parameter which may affect halo properties. 
Therefore, in this section, we study the effect of evolutionary stage of halos on 
the star-formation properties of halos. 
 
Observationally,  the magnitude gap between the most and second most bright 
member galaxies has widely been adopted as an indicator for the evolutionary 
status of the cluster. 
In this study, we use the stellar-mass ratio between them instead of the magnitude gap.
 
\begin{figure}
    \includegraphics[width=\columnwidth]{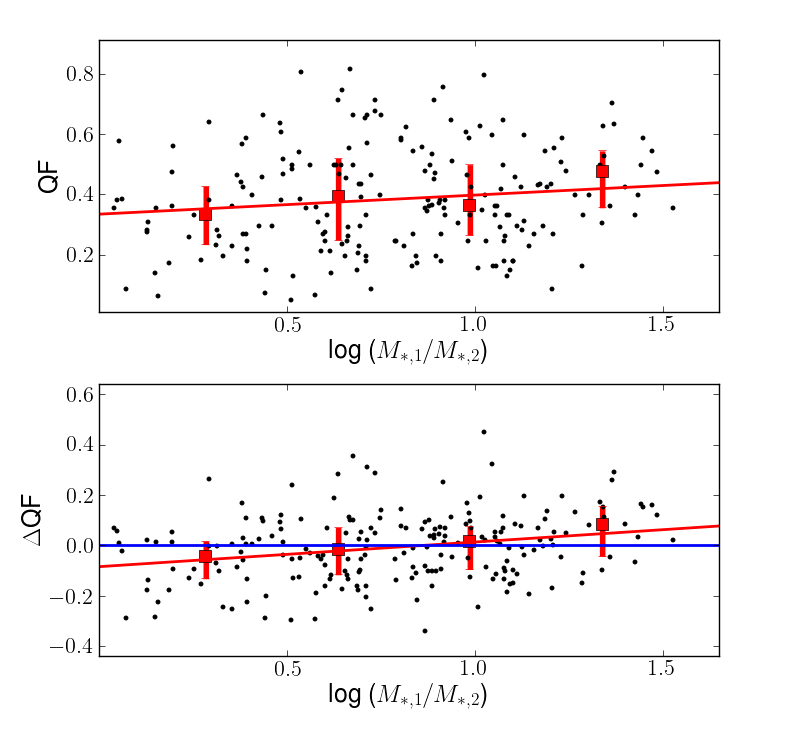}
    \caption{{\bf [Top]} Correlation between QF and the stellar-mass ratio of clusters and groups. 
Stellar mass ratio is defined as $M_{*,1}/M_{*,2}$, where $M_{*,1}$ and $M_{*,2}$ are the stellar 
masses of the most massive and second most massive galaxies within each halo's member galaxies. 
Red square symbols, along with their respective error bars, denote the median and NMAD of QF 
for each stellar-mass ratio bin. The red straight line shows the best-fit correlation. 
{\bf [Bottom]} Correlation between stellar-mass ratio and the residual QF, $\Delta$QF. 
Here, the residual QF, defined as (${\rm QF_{i} - QF_{HSFMS}}$), is the deviation from the best-fit HSFMS 
(as descibed in Equation~\ref{eq:qfm200}) for individual clusters and groups.}
    \label{fig:qfmsrat}
\end{figure}

Figure~\ref{fig:qfmsrat} depicts the correlation between QF and the stellar-mass ratio, $M_{*,1}/M_{*,2}$. 
As shown in the upper panel of this figure, galaxy groups or clusters with a higher steller-mass 
ratio tend to  have a higher QF. The best-fit line, represented by a red line, has a slope of 0.06, 
with a standard error of mean (SEM) of 0.012. 
This implies that there is a weak positive correlation between QF and the stellar-mass ratio in 
galaxy groups or clusters. 
The Spearman correlation coefficient, $\rho$, between QF and the stellar-mass ratio is 0.13 (with $p$-value of 0.068).

As seen in Section~\ref{sec:qfhmass}, there exists a correlation between QF and halo mass. 
We examine whether a correlation between the stellar-mass ratio and QF persisted even after 
removing the influence of halo mass on QF. 
To investigate this, we analyzed the relationship between the residual QF and the stellar-mass ratio, as depicted in the lower 
panel of the figure.
Here, the residual QF is defined as follows:
\begin{equation}
\Delta {\rm QF = QF_{i} - QF_{HSFMS}},
\label{eq:residqf}
\end{equation}
where ${\rm QF_i}$ and ${\rm QF_{HSFMS}}$ are QF of individual halo and QF 
on the best-fit HSFMS (Equation \ref{eq:qfm200}) for each halo with a given halo mass, respectiviely. 
As shown in this figure, there is a positive correlation between residual QF and stellar-mass ratio. 
The slope of the best-fit line, shown as a red line, is 0.10, and the Spearman correlation coefficient is 0.25 
(with $p$-value of 0.000), confirming the existence of a correlation between QF and the stellar-mass ratio, 
even when the impact of halo mass on QF is factored out.

%Figure~\ref{fig:qfmsrat} shows the correlation between QF, 
%stellar mass ratio, and halo mass. 
%As can be seen in this figure, galaxy groups or clusters with a larger steller-mass 
%ratio tend to  have a higher QF for a given halo mass. 
%This means there is a positive correlation between QF and stellar-mass ratio in 
%galaxy groups or clusters. 

%\begin{figure}
  %  \includegraphics[width=\columnwidth]{qf_dqf_msrat.png}
    %\caption{{\bf (Top)} Correlation between halo mass and stellar mass ratio of clusters and groups. 
%Stellar mass ratio is defined as $M_{*,1}/M_{*,2}$, where $M_{*,1}$ and $M_{*,2}$ are the stellar 
%masses of the most massive and the second most massive galaxies among the member galaxies in each halo. 
%Symbol colors are coded by QF. At a given halo mass, groups or clusters with larger stellar mass ratio tend to have 
%higher QF.}
   % \label{fig:qfmsrat}
%\end{figure}

The stellar mass of central galaxy (or BCGs) will keep increasing by 
merging satellite galaxies. 
Therefore, the stellar-mass difference becomes increasing with time unless 
a new massive galaxy enters into the cluster (as a member of merging group or cluster)
or the second most massive galaxy experiences the same amount of merging activities.  
The quiescent galaxy fraction also increases as time goes on, leading to 
the positive correlation between the quiescent fraction and the stellar-mass 
ratio as shown in Figure~\ref{fig:qfmsrat}. 
In other words, the evolutionary stage of clusters, parametrized with 
the stellar-mass ratio, can also affect the star-formation properties of clusters.  

\begin{figure}
    \includegraphics[width=\columnwidth]{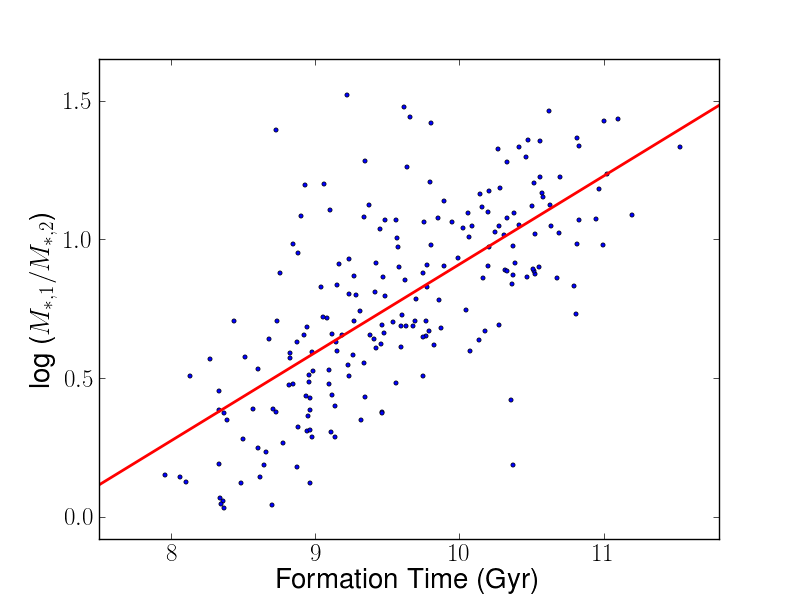}
    \caption{Correlation between the halo formation time and the stellar-mass ratio. 
The halo formatioin time is defined as the time at which half of the mass was assembled. 
The stellar-mass ratio exhibits a tight positive correlation with the halo formation time.}
    \label{fig:tfmsr}
\end{figure}

We examine the correlation between the stellar-mass ratio and formation time of galaxy clusters. 
The halo formation time is defined as the time at which half of the mass was assembled. 
As seen in Figure \ref{fig:tfmsr}, there exists a positive correlation between these two quantities, 
indicating that the stellar-mass ratio can serve as a good proxy for the evolutionary stage of galaxy clusters. 

\begin{figure}
    \includegraphics[width=\columnwidth]{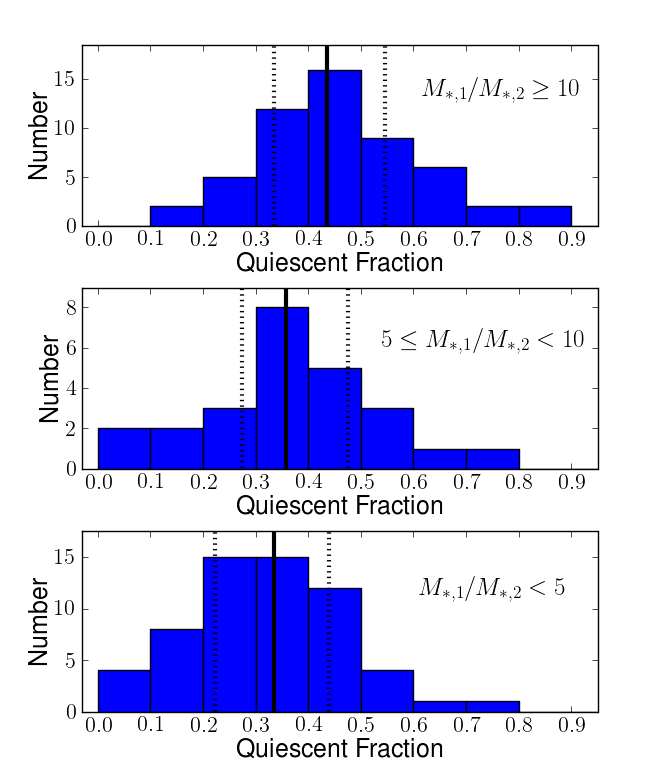}
    \caption{Quiescent galaxy fraction histograms for different stellar mass ratio for halos whose halo mass 
is within $1.5 \times 10^{13} M_{\odot} \le M_{200} < 4.0 \times 10^{13} M_{\odot}$ range.
{\bf [Bottom]} Halos with $M_{*,1}/M_{*,2} < 5$. {\bf [Middle]} Halos with $5 \le M_{*,1}/M_{*,2} < 10$.
{\bf [Top]} Halos with $M_{*,1}/M_{*,2} \ge 10$. In each panel, black solid and dotted vertical lines 
show the median and first and third quartile values of QF at each halo mass bin. This 
figure shows that QF is higher on average for halos with larger stellar mass ratios.}
    \label{fig:qfhistmsbn}
\end{figure}

This correlation between QF and stellar-mass ratio turns out to be 
clearer for intermediate halo-mass sample ($1.5 \times 10^{13} ~{\rm M_{\odot}} \le 
M_{200} < 4.0 \times 10^{13}~{\rm  M_{\odot}}$) as can be seen in 
Figure~\ref{fig:qfhistmsbn}. 
In this halo-mass range, the quiescent galaxy fractions of halos 
with lower stellar mass ratio (bottom panel) are smaller than the ones with 
higher stellar mass ratios (middle of top panel), reflecting the difference in 
their average evolutionary stage. 
QFs for more massive halos ($M_{200} \ge 4.0  \times 10^{13} ~{\rm M_{\odot}}$) 
are large regardless of their stellar-mass ratio, because the halo mass plays a
more dominant role in shaping the SF properties of clusters.

\begin{figure}
    \includegraphics[width=\columnwidth]{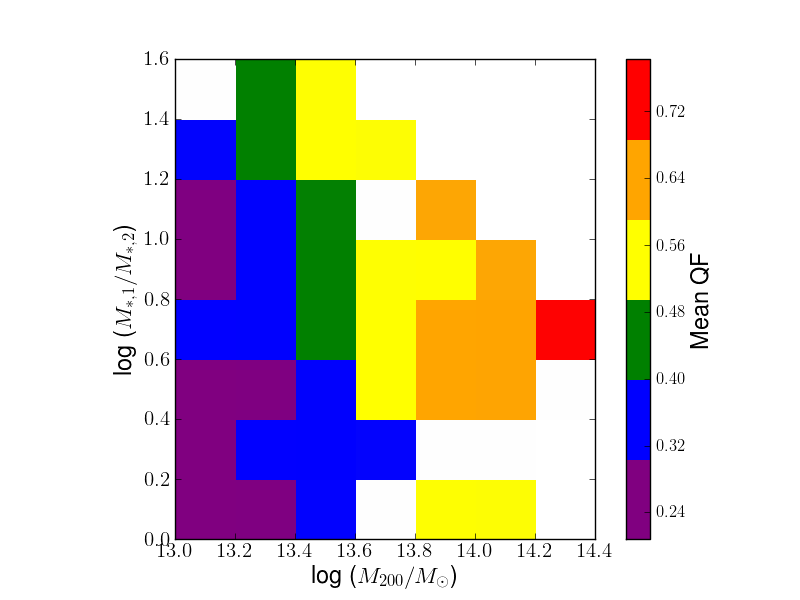}
    \caption{The correlation between halo mass (on the $x$-axis), stellar mass ratio (on the $y$-axis), 
and QF. We partition the parameter space of halo mass and stellar mass ratio into bins and 
compute the average QF for galaxy clusters or groups within each bin, with the results represented 
using colors. White-colored bins indicate the absence of galaxy clusters.}
    \label{fig:mhmsrqf}
\end{figure}

Figure~\ref{fig:mhmsrqf} depicts the correlation between halo-mass, stellar-mass ratio, 
and QF. 
In the parameter space defined by halo-mass and stellar-mass ratio bins, 
we calculate the average values of QF of galaxy clusters within each bin 
and represent them using colors. 
Upon examination of this figure, we can observe the following: 
(1) We confirm the relationship between halo-mass and QF, as described
in Section~\ref{sec:qfhmass}. 
As we move to the right in the figure, there is an increase in the 
prevalence of redder bins.
(2) We validate the relationship between stellar-mass ratio and QF, 
as discussed in Figure~\ref{fig:qfmsrat}. 
Moving downward in the figure, we see more blue and purple bins. 
This implies that galaxy groups or clusters with lower stellar-mass ratios 
within a given halo mass tend to exhibit lower QF values.
(3)  Galaxy clusters with greater halo mass exhibit a narrower range of stellar-mass ratios,
with a noticeable decrease in the maximum value of the stellar-mass ratio 
associated with these clusters. 

We perform a QF - halo mass - stellar-mass ratio multilinear fit to examine the extent 
to which the stellar-mass ratio contributes to the scatter in the QF - halo-mass correlation.
The mean square error (MSE) value obtained from this multilinear fit is approximately 7 \% 
smaller than the MSE from the QF - halo-mass linear fit, indicating that the stellar-mass ratio 
contributes to the scatter to some extent. 

\begin{figure*}
    \includegraphics[width=2.0\columnwidth]{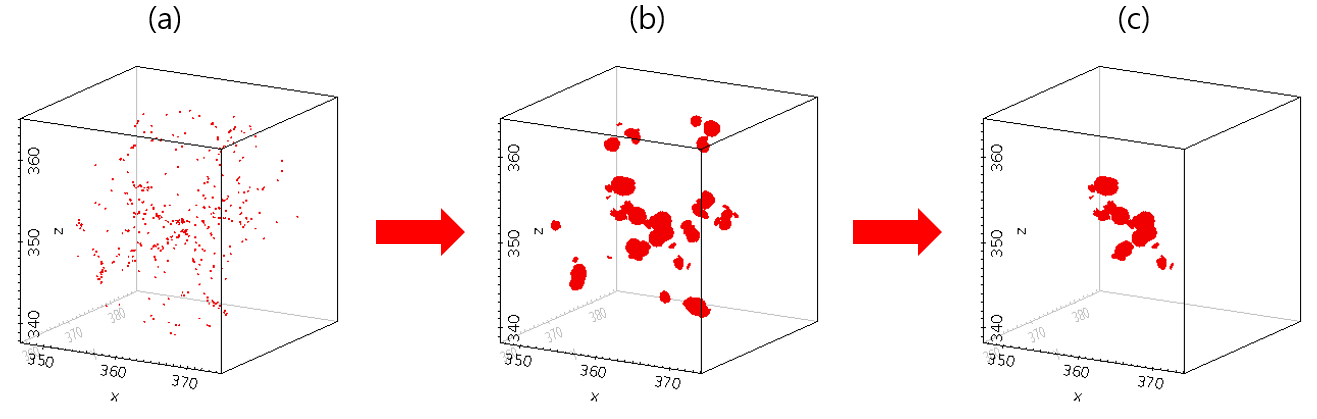}
     \caption{Procedure we used to measure the FoF fraction for an 
example halo. (a) Distribution of galaxies in and around the halo in three 
dimension. (b) Three dimensional distribution of pixels whose galaxy number density 
are greater than 1.4 $\sigma$. (c) Distribution of pixels which are connected to 
the central halo, among the pixels shown in (b), found by FoF algorithm.}
    \label{fig:fofex}
\end{figure*}

\subsection{Quiescent Galaxy Fraction and Large-scale Environment}
\label{sec:qflss}
%\subsubsection{Correlation between Quiescent Galaxy Fraction and FoF Fraction}
%\label{sec:qffof}

In the previous sections, we have found that the star-formation property of member galaxies 
is affected by the properties of their host halo, such as the halo mass and stellar mass ratio. 
Now, we turn our attention to the large-scale environment of halos. 

Previously, we found that there exists a correlation between the QF of 
high-redshift galaxy clusters and their large-scale environment, in 
a sense that clusters which are more connected to dense environment 
have lower level of quiescent galaxy fraction \citep{lee19}. 

According to \citet{lee19}, the large-scale environment is parameterized as 
the 'FoF fraction', which measures how much the fraction of the area 
surrounding the given cluster is occupied by dense structure with 
high galaxy number-density (above a given threshold). 
Therefore, when the cluster is embedded in a dense environment, the FoF fraction 
is higher. 
On the other hand, a relatively isolated cluster may have a lower FoF fraction. 

Here, we use the similar definition of FoF fraction as in \citet {lee19}, 
but we measure it in 3 dimensional space.  
 It is shown in \citet{lee19} that FoF fraction calculated in projected two dimensional 
space well correlates with FoF fraction measured in three dimension.

The FoF fraction in this work is calculated as follows: 
First, we place a sphere at the center of a given cluster with the radius of 5 Mpc. 
And we construct the galaxy number-density field.
Next, we find the overdense pixels by applying the density threshold of 1.4 $\sigma$ 
higher than the mean number density. 
Then, we link those overdense pixels using FoF algorithm with a linking length of 2 Mpc. 
Finally, the FoF fraction is defined as the volume ratio of the entire linked pixels to 
the 5 Mpc sphere.
Figure~\ref{fig:fofex} illustrates this procedure with a typical halo.
%We repeat this calculation changing the radius to 5 Mpc as well. 

\begin{figure}
    \includegraphics[width=\columnwidth]{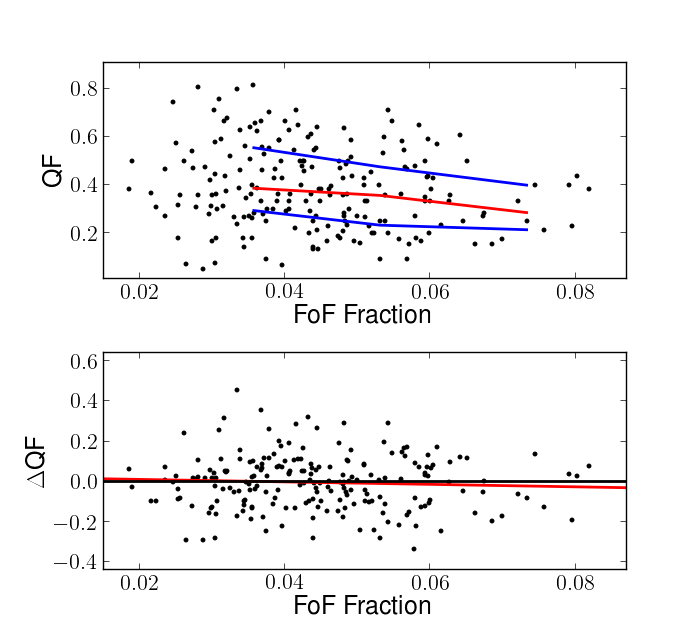}
     \caption{{\bf [Top]} Correlation between quiescent galaxy fraction and 
the FoF fraction. Red and blue lines show the median and the first and third quartile 
values of QF as a function of FoF fraction. Refer Section~\ref{sec:qflss} for the definition of the FoF fraction. 
 There is a weak anti-correlation, with a dearth of clusters with high FoF fraction and high QF. 
{\bf[Bottom]} $\Delta$QF (=${\rm QF_{i} - QF_{HSFMS}}$) as a function of FoF fraction. 
The red solid line represents the best linear fit to the  data. The scatter in QF decreases with increasing FoF fraction.}
    \label{fig:qffofv4} 
\end{figure}

The upper panel of Figure \ref{fig:qffofv4} depicts the correlation between the QF and the FoF fraction.
In this figure, we observe an anti-correlation between these two in line with 
our previous observational findings as reported in \citet{lee19}. 
In detail, when the FoF fraction is high (indicating that a cluster is embedded in 
dense environment), the QF value tends to be low. 
Every cluster with the FoF fraction $>0.07$ has the QF value smaller than 0.5. 
On the other hand, relatively isolated clusters with a low FoF 
fraction exhibit a broader range of QF values.
This is because when a cluster is relatively isolated or well centrally concentrated 
(with low FoF fraction), 
other various factors (as seen in the previous sections) affect the QF values. 
On the other hand, if a cluster has an extended distribution of galaxies and with 
a higher probability, therefore, it is more connected with (or embedded within) 
dense environment, the QF values are kept low due to the infalling material. 

It is worth noting that the strength of this correlation is somewhat weaker 
compared to what we observed in observation. 
The Spearman correlation coefficient is -0.16 (with $p$-value of 0.018), indicating 
a weak anti-correlation.
We speculate that this is because the SF quenching in HR5 galaxies may not 
be rapid enough (or delayed  due to relatively weak AGN feedback) 
at this redshift.

The bottom panel of Figure \ref{fig:qffofv4} illustrates the relationship between the FoF fraction 
and  $\Delta$QF (=${\rm QF_{i} - QF_{HSFMS}}$). 
A anti-correlation is observed between these two variables, 
as indicated by the red solid line, which has a slope of -0.62 (with a SEM of 0.0093).

%We can confirm the dependence of QF on the FoF fraction in Figure~\ref{fig:qfhstfofbin}, 
%which shows the QF distribution for halos with different FoF fractions. 
%The top panel of this figure shows the QF distribution for halos with FoF fraction 
%smaller than 0.04, i.e. relatively isolated halos. 
%The QF values of these isolated halos are on average higher than the halos with more 
%higher value FoF fractions (middle panel: $0.045 \le$ FoF fraction $\le 0.065$, top panel: 
%FoF fraction $\ge 0.07$).

%After dividing our sample based on halo mass, we examine the relationship 
%between QF and FoF fraction (Figure~\ref{fig:qffofmhbin}. 
%As seen in this figure, the (aniti-)correlation  between QF and FoF fraction is more 
%evident in low and intermediate halo-mass bins (left and middle panels in the figure), while 
%the correlation is not observed in the most massive bin (right panel in the figure). 
%The majority of massive clusters, with log (M$_{h}/M_{\odot}) > 13.5$, exhibit 
%a low FoF fraction. 
%
To summarize, we find the large-scale structure, measured as FoF fraction 
is another important factor affecting on the SF property of clusters, 
confirming our previous finding in \citet{lee19}.

\section{Discussion: Why do more massive halos show higher QF?}
\label{sec:disc}

%\subsection{Why do more massive halos show higher QF?}
%\label{sec:disc_hsfms}

In Section~\ref{sec:qfhmass}, we demonstrate that QFs of more massive halos 
are higher --- i.e., larger fraction of member galaxies are quiescent 
in more massive halos
This aligns with previous observational results of \citet{kim09,woo13}. 
Then we can ask what makes this correlation between halo mass and QF. 
Satellite galaxies in more massive halos may experience stronger ram pressure 
stripping due to higher peculiar velocity \citep[e.g.][]{bek09}. 
Another possibility is that more massive halos may experienced more mergers 
with or accretion of 
groups or clusters. The satellite galaxies entered massive halos through this accretion 
of groups or clusters may be quenched in these groups or clusters 
\citep[i.e., pre-processing][]{bia18}. 
Knowing which, among these, acts dominantly in shaping the correlation between 
QF and halo mass is out of scope of this paper. 
We plan to address this issue in our future work. 

In the QF-halo mass diagram (Figure~\ref{fig:qfmh}), perhaps more important 
than the correlation between QF and halo-mass is the existence of a lower envelope, 
indicating the presence of a minimum QF value that galaxy clusters have at a given halo mass. 
This suggests that in more massive halos, the quenching of member galaxies may occur more 
rapidly, or that these massive halos may impede the inflow of new star-forming galaxies.
A more in-depth analysis of this phenomenon may be possible by 
studying how the QF-halo mass relationship evolves across a broader range of redshifts. 
We leave this as our future work.

\section{Conclusion}
\label{sec:conc}

Using model galaxies and halos at $z=0.873$ from Horizon Run 5 simulation, 
we investigate the star formation properties of galaxy clusters and groups 
and their dependence on various internal and external properties of these 
groups and  clusters. 

%\begin{figure}
  %  \includegraphics[width=\columnwidth]{qf_fof_lmhbin_up.png}
    % \caption{Correlation between quiescent galaxy fraction and the FoF fraction at three 
  %   halo-mass bins. Halo-mass range in each bin is [log (M$_{h}/M_{\odot}) \leq 13.3$; 
%{\bf left}], [$13.3 < $log (M$_{h}/M_{\odot}) \leq 13.5$; {\bf middle}], and 
%[log (M$_{h}/M_{\odot}) > 13.5$; {\bf right}]. The correlation between QF and FoF 
%is most pronounced within the lowest halo-mass bin, whereas it appears notably weaker 
%within the highest mass bin.}
 %   \label{fig:qffofmhbin}
%\end{figure}
%
From this investigation, we found several interesting results, as follows: 

\begin{enumerate}

\item [1.] The quiescent galaxy fraction of clusters and groups show a 
positive correlation with their halo mass, $M_{200}$, which we dub the
`halo star-formation main sequence (HSFMS)'. 
We also find the halo-mass-dependent minimum values 
in quiescent galaxy fraction. 
Combined together, these results show that the halo mass is an 
important parameter to determine the quiescent galaxy fraction of 
clusters: majority of member galaxies in massive clusters are quiescent 
at $z \sim 0.87$ already, while quiescent galaxy fraction spans broader range 
for less massive clusters or groups. 
The quiescent galaxy fraction of low mass halos 
is further affected by several other factors.

\item [2.] We find that there exists a positive correlation between the quiescent 
galaxy fraction and the stellar mass ratio in a sense that halos with larger 
stellar mass ratio between the most and the second most member galaxies 
tend to have higher QF values. 
This trend is clearer for intermediate mass halos 
($1.5 \times 10^{13} M_{\odot} \le M_{200} < 4.0 \times 10^{13} M_{\odot}$). 
This stellar-mass dependence indicates that the evolutionary stage of a 
halo also affects the star-formation property of clusters or groups, 
especially the intermediate-mass halos.  
This stellar-mass ratio dependence of QF contributes to the scatter in QF for given 
halo mass, which is larger for intermediate mass halos.

\item [3.] We investigate the dependence of the QF of halos on the 
large-scale environment, measured by FoF fraction. We find a weak correlation 
between these two properties, in a sense that clusters or groups which 
have more connected nearby overdense large-scale structure tend to 
have lower values of QF. 
This result confirms our previous, observational finding \citep{lee19}. 

%\item [4.] There also exist a weak anti-correlation between the QF and the 
%nearest neighbor distance for sub-samples of intermediate-mass halos, 
%upto $\sim  18$ normalised nearest neighbor distance only for halos 
%with the 'PS'-type nearest neighbor. 
%Specifically, the QF of a halo increases as the nearest neighbor 
%distance decreases only when the neighbor is 'PS'-type. 
%This trend is also more clearly seen for intermediate mass halos.
%We interpret this trend as a halo version of conformity signal. 

\end{enumerate}

In this study, we delve into the SF characteristics of galaxy clusters 
and groups at a redshift of $z=0.873$, with a specific focus on identifying 
the key factors that influence these properties. 
Intriguingly, our research reveals that multiple parameters, such as 
halo-mass, the evolutionary stage of the halo, and the large-scale environmental context, 
collaborate in shaping the SF status of clusters and groups at this specific redshift. 
Expanding our investigation to encompass a wider range of redshifts could provide 
a more comprehensive comprehension of the evolutionary trends governing 
these aspects of clusters and groups.

\section*{Acknowledgements}

The authors thank the Korea Institute for Advanced Study for providing 
computing resources (KIAS Center for Advanced Computation Linux Cluster 
System) for this work. This work was supported by the National Research 
Foundation of Korea (NRF) grant funded by the Korean government 
(MSIT; Nos. 2020R1I1A1A01060310, 2020R1A2C3011091, 2021M3F7A1084525). 
JK was supported by a 
KIAS Individual Grant (KG039603) via the Center for Advanced Computation at 
Korea Institute for Advanced Study. This work benefited from the 
outstanding support provided by the KISTI National Supercomputing Center 
and its Nurion Supercomputer through the Grand Challenge Program 
(KSC-2018-CHA-0003 and KSC-2021-CHA-0012). Large data transfer was supported 
by KREONET, which is managed and operated by KISTI. 
JL is supported by the National Research Foundation of Korea (NRF-2021R1C1C2011626).  
BKG and CGF acknowledge the support of the European Union's Horizon 
2020 Research and Innovation Programme (ChETEC-INFRA - Project no. 101008324) 
and ongoing access to viper, the University of Hull's High Performance Computing Facility.
YK is supported by Korea Institute of Science and Technology Information (KISTI) 
under the institutional R\&D project (K24L2M1C4). 
This work was also partly supported by the National Research Foundation of Korea(NRF) 
grant funded by the Korea government (MSIT, 2022M3K3A1093827).

%%%%%%%%%%%%%%%%%%%%%%%%%%%%%%%%%%%%%%%%%%%%%%%%%%
\section*{Data Availability}

HR5 data are not publically available, unless a researcher is in collaboration with HR5 team members. 
We encourage any researcher who is interested in using the HR5 to contact Prof. Changbom Park (cbp@kias.re.kr).

%The inclusion of a Data Availability Statement is a requirement for articles published in MNRAS. Data Availability Statements provide a standardised format for %readers to understand the availability of data underlying the research results described in the article. The statement may refer to original data generated in the %course of the study or to third-party data analysed in the article. The statement should describe and provide means of access, where possible, by linking to the %data or providing the required accession numbers for the relevant databases or DOIs.



\bibliographystyle{mnras}
\bibliography{hr5ref} % if your bibtex file is called example.bib

%\begin{thebibliography}{99}
%\end{thebibliography}

%%%%%%%%%%%%%%%%%%%%%%%%%%%%%%%%%%%%%%%%%%%%%%%%%%

%%%%%%%%%%%%%%%%% APPENDICES %%%%%%%%%%%%%%%%%%%%%

%%\appendix

%%\section{Some extra material}

%%If you want to present additional material which would interrupt the flow of the main paper,
%%it can be placed in an Appendix which appears after the list of references.

%%%%%%%%%%%%%%%%%%%%%%%%%%%%%%%%%%%%%%%%%%%%%%%%%%

% Don't change these lines
\bsp	% typesetting comment
\label{lastpage}
\end{document}